\documentclass{article}

\usepackage{arxiv}
\usepackage{url}
\usepackage{xcolor}
\usepackage{algorithm} 
\usepackage{stfloats}
\usepackage{algpseudocode} 
\usepackage[utf8]{inputenc} 
\usepackage[T1]{fontenc}    
\usepackage{hyperref}       
\usepackage{url}            
\usepackage{booktabs}       
\usepackage{amsfonts}       
\usepackage{nicefrac}       
\usepackage{microtype}      
\usepackage{lipsum}		
\usepackage{graphicx}
\usepackage{natbib}
\usepackage{doi}
\usepackage{multirow}
\usepackage{xcolor}
\usepackage{geometry} 
\usepackage{amsmath}
\usepackage{amssymb}
\usepackage{hyperref}
\usepackage{booktabs}
\usepackage{wrapfig}
\usepackage{longtable} 
\usepackage{caption}

\title{Aligning Large Language Models with Healthcare Stakeholders: A Pathway to Trustworthy AI Integration}

\author{
\textbf{Kexin Ding}$^{1}$, 
\textbf{Mu Zhou}$^{1}$, 
\textbf{Akshay Chaudhari}$^{2}$,
\textbf{Shaoting Zhang}$^{3,\dag}$,
\textbf{Dimitris Metaxas}$^{1,\dag}$\\
\\
$^1$Department of Computer Science, Rutgers University, New Jersey, USA \\
$^2$Department of Radiology, Stanford University, California, USA \\
$^3$Shanghai Artificial Intelligence Laboratory, Shanghai, China
}

\renewcommand{\headeright}{}
\renewcommand{\undertitle}{}

\hypersetup{
pdftitle={A template for the arxiv style},
pdfsubject={q-bio.NC, q-bio.QM},
pdfauthor={David S.~Hippocampus, Elias D.~Striatum},
pdfkeywords={Collaborative learning, Medical image, foundation model}
}

\begin{document}
\maketitle
\def\thefootnote{†}\footnotetext{Corresponding author, email: dnm@cs.rutgers.edu}

\begin{abstract}
The wide exploration of large language models (LLMs) raises the awareness of alignment between healthcare stakeholder preferences and model outputs. This alignment becomes a crucial foundation to empower the healthcare workflow effectively, safely, and responsibly. Yet the varying behaviors of LLMs may not always match with healthcare stakeholders’ knowledge, demands, and values. To enable a human-AI alignment, healthcare stakeholders will need to perform essential roles in guiding and enhancing the performance of LLMs. Human professionals must participate in the entire life cycle of adopting LLM in healthcare, including training data curation, model training, and inference. In this review, we discuss the approaches, tools, and applications of alignments between healthcare stakeholders and LLMs. We demonstrate that LLMs can better follow human values by properly enhancing healthcare knowledge integration, task understanding, and human guidance. We provide outlooks on enhancing the alignment between humans and LLMs to build trustworthy real-world healthcare applications. 
\end{abstract}

\section{Introduction}

Large Language Models (LLMs) represent a key effort in the generative AI wave, leveraging the power of extensive human knowledge and feedback \cite{ref1,ref2}. As they continue to evolve, LLMs could bring human-like reasoning capability into a range of healthcare scenarios for clinicians, patients, educators, and payers \cite{ref3,ref4,ref5,ref6} (as shown in Figure 1). The proper alignment between human and LLM becomes increasingly important, requiring LLMs’ outputs to match human preferences to safely assist clinical routines. Yet this is notoriously challenging in the healthcare system given the privacy restrictions and various clinical demands. Patients expect to deepen their understanding of disease management and prevention, while payers focus on evaluating patient clinical records before authorizing patients' insurance applications. 

Human-in-the-loop efforts are vital to improve the alignment between LLMs and healthcare stakeholders throughout the LLM development cycle, spanning pre-, intermediate-, and post-training, and inference stages. In the pretraining stage, LLMs require a large amount of domain-specific data to build the fundamental concepts of healthcare scenarios. Extra human effort in data labeling and quality verification is crucial to avoid poisoning content being learned and generated by LLMs. In the intermediate-training stage, instruction learning strategies are instrumental to allow LLMs to integrate up-to-date, domain-specific knowledge, such as patient record understanding \cite{ref7} and status forecast \cite{ref8}. With human-designed instructions, these LLMs could better generalize on a variety of healthcare tasks while flexibly producing human-desired outputs. In the post-training stage, human contribution is typically realized via Reinforcement Learning from Human Feedback (RLHF) \cite{ref9,ref10} toward the human-LLM alignment by guiding model outputs using human feedbacks.

To expedite the human-AI alignment in the inference stage, constructing a plain natural language instruction, also known as “prompt engineering”, becomes a viable path to enable LLM inference to generate relevant answers to various healthcare queries. For instance, chain-of-thought (CoT) \cite{ref11} and self-consistency prompting \cite{ref12} are growingly used to guide LLM in producing a more reliable response to user queries. Guided by a human-designed reasoning chain, LLMs can produce a step-by-step path to the output, enabling a better alignment performance and mitigating fictitious contents. Instead of simply providing diagnosis suggestions, using CoT technology can provide a step-by-step reasoning interpretation by mimicking the clinician's thoughts and analysis process \cite{ref13}. 

These human-LLM interactions continue to evolve in terms of aligning human values and model behaviors on a topic of interest \cite{ref14}. In this review, we discuss the critical alignment between healthcare stakeholders and LLMs in four critical scenarios, including clinical workflow, patient care, medical education, and healthcare payers. We provide the outlook of safe alignment of LLMs, encompassing key concerns on data distribution, model development reliability, and LLM regulation in a broad landscape of healthcare applications.

\begin{figure}[H]
    \centering
    \includegraphics[width=0.8\textwidth]{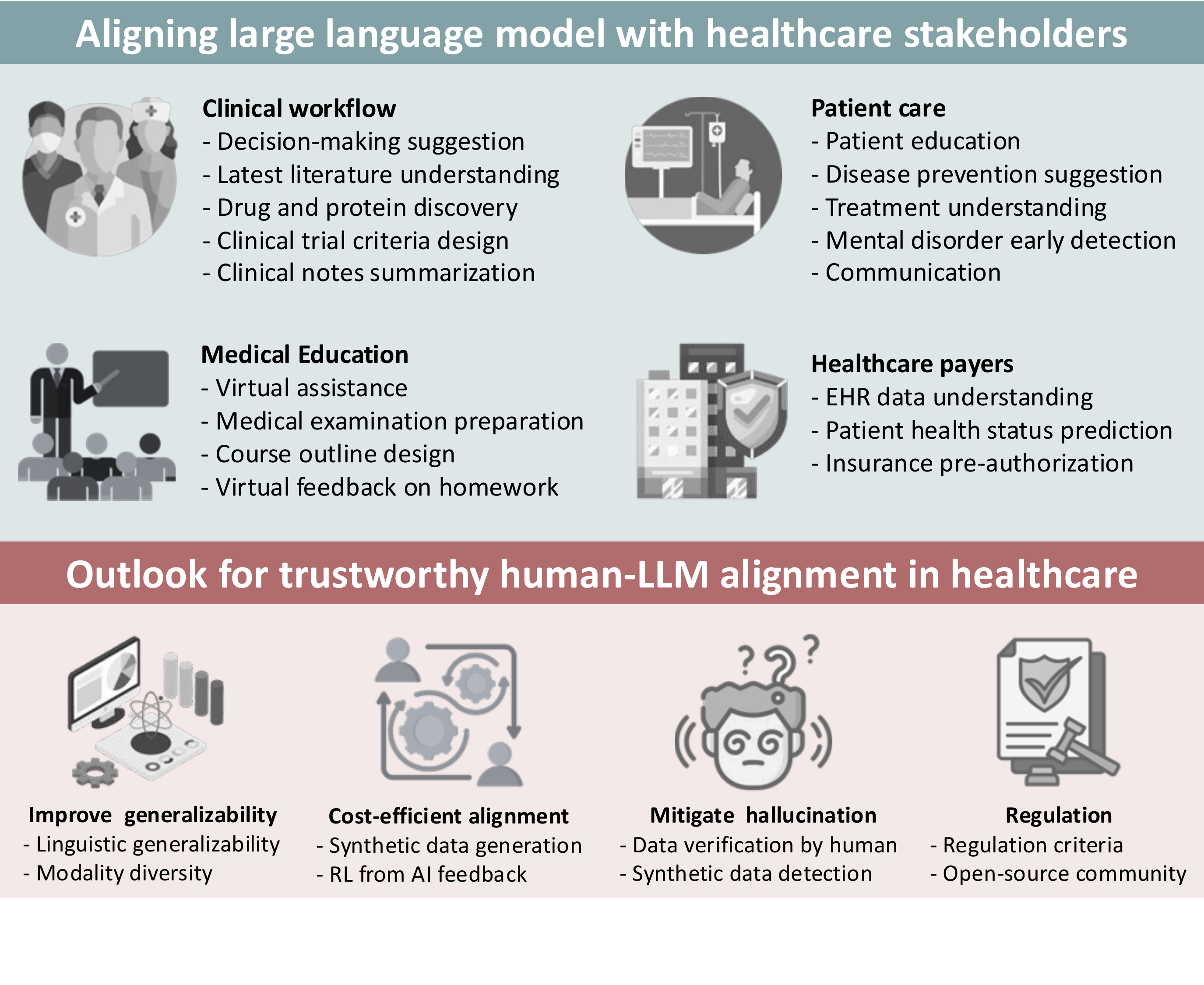} 
    \vspace{-6mm}
    \caption{Overview of aligning large language models (LLMs) with key healthcare stakeholders and the outlook for trustworthy human–LLM collaboration. We highlight major alignments between healthcare stakeholders and LLMs across various scenarios from clinical workflow, patient care, medical education, and healthcare payers. We also outline future directions for trustworthy human-LLM alignment in healthcare, including improving generalizability, developing cost-efficient alignment, mitigating hallucinations, and establishing regulatory frameworks.}
    \label{fig:llm-healthcare}
\end{figure}

\section{Large Language Model Alignment}
Large language models (LLMs) have excelled at natural language processing (NLP) in conversation, translation, and summarization tasks, exemplified by ChatGPT \cite{ref15}, LlaMA \cite{ref1}, PaLM \cite{ref16}, and Falcon \cite{ref2}. A key challenge of adopting LLMs into healthcare is that the training objective of general-domain LLMs (e.g., next-word prediction) does not reflect the complex scenarios in clinical laboratories, hospitals, and payers \cite{ref17}. Knowledge integration and task understanding are two hurdles causing the hallucination phenomena when applying LLMs in healthcare. Misleading patient disease understanding and physicians' treatment decision-making are likely seen when LLM generates incorrect suggestions for critical queries of diseases. To address this dilemma, guiding LLM behavior in matching human values becomes increasingly crucial \cite{ref18}. Unlike traditional language tasks (e.g., word prediction), solving healthcare tasks is more complex because of the necessity of reliable reasoning capabilities and domain knowledge. Enhancing LLM capability in capturing healthcare data patterns is critical for the alignment in healthcare \cite{ref19}. In the following, we discuss major techniques to align human knowledge and LLMs to enhance human-AI interactions and mitigate the misleading information produced by LLMs.

In the pretraining and intermediate training stage, LLM parameters are updated with healthcare-specific training data \cite{ref20}, while in the post-training, LLMs are guided directly by human feedbacks to ensure LLMs’ output is more aligned with healthcare stakeholder’s preferences. In the pretraining stage, PubMedBERT \cite{ref21} and GatorTron \cite{ref7} are examples of training LLMs from scratch with healthcare vocabulary from large-scale biomedicine article publications. Considering the daunting challenge of large-scale healthcare data curation for training LLMs from scratch, domain-specific pretraining \cite{ref22} is recommended for aligning general-purpose LLMs in the healthcare domain. Such pretraining mechanisms could integrate healthcare patterns while utilizing the powerful generalizability of general-domain pretrained LLMs. For instance, sciBERT \cite{ref23}, ClinicalBERT \cite{ref8}, and BioBERT \cite{ref24} all integrate solid healthcare knowledge into the general-purpose BERT \cite{ref25}. To improve the efficiency, in the intermediate-training stage, data- and parameter-efficient finetuning (PEFT) can help align general-domain LLMs and healthcare stakeholders while mitigating model catastrophic forgetting. Available tools from prompt tuning, prefix tuning, and Low-Rank Adaptation (LoRA) \cite{ref26} can help LLMs accomplish healthcare tasks at lower training costs and human efforts. For instance, LoRA has been widely known as an efficient adaptor for pretraining open-source LlaMa into clinical note understanding \cite{ref11}. Instruction-based alignment \cite{ref27} is another useful tool to address the misalignment between the general LLMs and healthcare tasks with minimal human effort. To illustrate, human experts provide well-designed instructions to guide the LLMs to meet human expectations \cite{ref9}. Based on pretrained LlaMA-2 \cite{ref28}, MEDITRON-70B \cite{ref29} is built with medical data from PubMed and task-specific instructions. Similarly, PMC-LLaMA \cite{ref30} and Radiology-Llama2 \cite{ref27} use instruction tuning to adapt the original LLM into healthcare with biomedical publications and medical textbooks. In the post-training stage, to better exploit human feedback on model outputs, leading studies \cite{ref10,ref31,ref32} have demonstrated that human experts could use Reinforcement Learning from Human Feedback (RLHF) to generate an LLM response that better matches human preferences by training a reward model guide LLM with the human feedback \cite{ref9,ref10}.

In the inference stage, emerging prompt engineering is a process to inject human guidance into model inference and reasoning, allowing medical centers to access the specified healthcare information for their queries without sharing raw patient data. As a leading prompt-based technique, Chain-of-thought (CoT) designs a step-by-step reasoning chain as a prompt to enhance the reasoning ability of LLMs. With human-designed prompts for revealing the thought processes, LLMs can generate reasoning steps to enhance the cognition alignment between humans and models. Google developed Med-PaLM 2, which utilized chain-of-thought (CoT) and self-consistency \cite{ref12} techniques to provide a clear reasoning chain for finding reliable solutions (i.e., the model scored up to 86.5\% on the MedQA dataset and improved upon Med-PaLM by over 19\% \cite{ref33}).

\section{Alignments between Stakeholders and LLM in Healthcare}
We discuss the alignment between healthcare stakeholders and LLMs in real-world scenarios, from clinical workflow, patient care, and medical education to health insurance. We provide a brief summarization of alignment scenarios between LLM and healthcare stakeholders in Table 1.

\renewcommand{\arraystretch}{1.5}

\begin{longtable}{|p{3.2cm}|p{3.2cm}|p{8.5cm}|}
\hline
\textbf{Study} & \textbf{Healthcare scenario} & \textbf{How does LLM align with human demand} \\
\hline
\endfirsthead

\cite{ref34,ref35,ref11} & Clinical routine & Improving work productivity in clinical decision-making. \\
\hline
\cite{ref13} & Clinical routine & Using diagnostic reasoning prompts to help with diagnosis and interpretation. \\
\hline
\cite{ref39} & Clinical routine & Using ChatGPT to generate diagnosis notes. \\
\hline
\cite{ref3} & Clinical routine & Accelerating timely scientific finding integration in treatment decision-making. \\
\hline
\cite{ref40} & Clinical routine & Clinical record summarization. \\
\hline
\cite{ref34} & Clinical routine & Clinical letter generation. \\
\hline
\cite{ref35} & Clinical routine & Rewriting clinical vignettes with caveats and review. \\
\hline
\cite{ref41} & Protein exploration & Generating protein sequences. \\
\hline
\cite{ref42} & Protein exploration & Translating protein sequences by natural language to accelerate protein annotation and function description. \\
\hline
\cite{ref43} & Gene exploration & Gene-phenotype analysis. \\
\hline
\cite{ref44} & Gene exploration & Genetic biomarker hypothesis generation. \\
\hline
\cite{ref45} & Drug discovery & Designing and optimizing the pharmacokinetics and pharmacodynamics of new drugs. \\
\hline
\cite{ref46} & Clinical trial & Simplifying the eligibility criteria design. \\
\hline
\cite{ref47} & Clinical trial & Interpretable criterion eligibility generation and trial matching based on patient notes. \\
\hline
\cite{ref48} & Clinical trial & Patient pre-screening by determining patient eligibility. \\
\hline
\cite{ref49} & Clinical trial & Clinical trials summarization related to wearable devices. \\
\hline
\cite{ref4} & Patient Education and communication & Providing recovery and preventional suggestions to patients who are recovering from severe COVID-19. Providing patient-friendly questions to guide patients in explaining concerns and confusions. \\
\hline
\cite{ref53} & Patient Education & Supporting maritime health for seafarers on long voyages at sea. \\
\hline
\cite{ref52} & Patient Education & Drug-drug interaction interpretation. \\
\hline
\cite{ref54} & Patient mental health detection & Promoting timely detection of various mental disorders (e.g., depression, anxiety, and suicidal ideation) from patients. \\
\hline
\cite{ref55} & Patient mental health detection & Achieving clinical contextual knowledge for identifying major depressive disorder phenotypes from clinical notes. \\
\hline
\cite{ref50} & Patient care & Patient caregivers can receive organized psychological and practical recommendations for patient care. \\
\hline
\cite{ref36,ref37,ref38} & Medical Education & Demonstrate that ChatGPT can pass medical professional exams. \\
\hline
\cite{ref59} & Medical Education & Drafting course content outlines and creating assessment questions with minor human revisions. \\
\hline
\cite{ref5} & Medical Education & Student performance feedback generation that coherently summarizes students’ performance. \\
\hline
\cite{ref60} & Medical Education & Reformatting unstructured medical curriculum text, question-answering interactions, and clarifying lecture uncertainties. \\
\hline
\cite{ref61} & Health insurance administration & Writing pre-authorization requests for insurance companies. \\
\hline
\cite{ref7} & EHR understanding & Extracting and interpreting patient characteristics from EHRs. \\
\hline
\cite{ref8} & Medical event forecast & Predicting 30-day hospital readmission at various time points of admission, including early stages and discharge. \\
\hline
\cite{ref62} & Medical event forecast & Convert EHR document text into structured concepts for patient medical events forecasts. \\
\hline
\caption{Summarization of alignments between Stakeholders and LLM in Healthcare.}
\end{longtable}

\subsection{Clinical Workflow}
In the clinical workflow, physicians suffer from a heavy burden of routine operational tasks. With strong reasoning and summarization capabilities, LLMs can assist humans to improve work productivity in clinical decision-making \cite{ref34,ref35,ref11}. By using human-designed prompts, LLM can produce outcomes by reasonably mimicking the clinical reasoning processes of clinicians \cite{ref13}. Such reasoning chains could provide an interpretable rationale for clinicians to determine the reliability of LLM's outputs. Growing studies reveal that LLMs can achieve passing scores in medical qualification examinations \cite{ref36,ref37,ref38}, offering opportunities to generate diagnostic notes and accelerate clinical workflow operations \cite{ref39}. In addition, to support clinical interventions, timely scientific findings must be integrated into real-world clinical decision-making. Given the complex volumes of research articles, LLMs are useful tools to gather the latest insights in the literature. For instance, by aligning the novel scientific findings from the growing COVID literature, Llama model \cite{ref1} enables clinicians to incorporate timely knowledge in their clinical treatment decision-making \cite{ref3}. It is increasingly clear that LLM-enabled document analysis can play a key role in clinical record summarization \cite{ref40}, clinical letter generation \cite{ref34}, and rewriting clinical vignettes with caveats and appropriate review \cite{ref35}. 

Aligning LLMs with clinical laboratory data analytics (e.g., protein or gene sequencing) will open a wide scope of biological discoveries. To acquire insights from natural protein sequencing, a protein generation large language model (ProGen) \cite{ref41} is developed on a public dataset containing 280M protein sequences. After training, ProGen can be prompted to generate full-length protein sequences for any protein family from scratch. Yet such an alignment strategy heavily relies on data curation quality and computational resources. The prompt-augmented pretraining data allows Galactica to achieve promising inference capabilities. For example, Galactica shows good translation capabilities from protein sequences toward natural language, enabling the acceleration of protein annotation and function description \cite{ref42}. In addition, Med-PaLM2 \cite{ref43} applies an instruction finetuning to efficiently align with medical domain-specific knowledge to analyze gene-phenotype relationships and generate novel hypotheses towards genetic biomarker discoveries \cite{ref44}. To alleviate the burden on data curation, prompt engineering essentially imposes LLMs to provide human-expected information without the need to collect new data. The well-designed prompts can provide differential guidance for LLMs when aligning with human demand in designing and optimizing the pharmacokinetics and pharmacodynamics of new drugs \cite{ref45}. 

The strong reasoning capability of LLMs can be used to improve the efficiency of clinical trial workflow. AutoTrial \cite{ref46} simplifies the eligibility criteria design by pretraining on a large corpus of clinical trial documents. To better reflect human needs, AutoTrial is trained to generate specific criteria according to the input instructions by leveraging multi-step reasoning and mimicking the retrieved clinical eligibility criteria. To accelerate patient-trial matching, TrialGPT \cite{ref47} is able to produce interpretable criterion eligibility and select the proper trials based on patient notes. By leveraging the promising inference capability of LLMs, physicians can reduce their workload of patient pre-screening by determining patient eligibility for clinical trials \cite{ref48}. Similarly, CliniDigest \cite{ref49} is useful for providing truthful summaries of clinical trials related to wearable devices. It is remarkable that CliniDigest can reduce up to 85 clinical trial descriptions (approximately 10,500 words) into a concise 200-word summary with references.

\subsection{Patient Care}
Patients often have insufficient knowledge of managing complications of diseases \cite{ref50}. The advent of LLMs can match the demand from patients by providing relevant answers to diverse medical queries \cite{ref51}. Performing as a conversational encyclopedia, LLM can help patients acquire educational resources on disease knowledge, including disease prevention strategies, symptoms, treatment options, and post-treatment care \cite{ref4,ref50}. For example, ChatGPT provides recovery and preventional suggestions to patients who are recovering from severe COVID-19 \cite{ref4}. To understand the therapeutic risk, patients can understand complex drug-drug interactions by asking ChatGPT, “When or why can I take X and Y together?” (e.g., X and Y are two drug names) \cite{ref52}. Despite showing a high-level relevance, the answer quality can be improved by aligning LLMs with more scientific findings. Especially for patients who struggle to acquire healthcare information in a poor internet environment, such as seafarers on long voyages at sea \cite{ref53}, LLM can serve as a local knowledge assistant to provide valuable support for their health and welfare through medical consultation. 

Patients are highly vulnerable during treatment because of the mental and economic pressure. The demand for mental health detection and recovery support is rising among patients and their caregivers \cite{ref50,ref54}. LLMs could automatically detect the mental abnormality of patients and assist caregivers in providing timely emotional support. MentalBERT and MentalRoBERTa \cite{ref54} are examples of promoting timely detection of various mental disorders (e.g., depression, anxiety, and suicidal ideation) from patients. Similarly, BioClinical BERT \cite{ref55} achieves clinical contextual knowledge for identifying major depressive disorder phenotypes from clinical notes. In addition, patient caregivers can receive organized psychological and practical recommendations from ChatGPT, enabling better care of patients \cite{ref50}. Standing in the position of patients who are not well-educated with medical knowledge, ChatGPT, as an example, has assisted clinicians in asking patient-friendly questions, leading patients to clearly explain their concerns and confusions \cite{ref4}.

\subsection{Medical Education}
Medical education requires a high degree of factual precision to avoid misleading or inaccurate information delivery among educators and students \cite{ref56}. Meanwhile, educational participants have shown increasing interest in optimizing educational workflow using LLMs. To align with this goal, LLMs should first demonstrate the equivalent capability of human experts who have finished long-term studies with comprehensive medical knowledge. As evaluated by the United States Medical Licensing Examination (USMLE), covering topics from basic science, clinical reasoning, and medical management to bioethics \cite{ref37}, LLMs have shown a promising level of healthcare knowledge to pass professional examinations as a 3rd-year medical student \cite{ref36,ref37,ref38}. Such model cognition verification can help students and educators trust LLMs and get a handy response to their queries about professional medicine knowledge \cite{ref57}. To further help human professionals, using chain-of-thought technologies could advance LLM performance on complex tasks in medical professional exam preparations \cite{ref58}. 

In medical school, both educators and students have a heavy burden on their daily routines. Educators must invest significant time in course design and knowledge delivery, while students often struggle to understand and retain large volumes of complex information \cite{ref56}. LLMs provide a remarkable opportunity to assist both educators and students in enhancing the efficiency of medical education. To accelerate the teaching preparation, ChatGPT can draft course content outlines and create assessment questions with minor human revisions \cite{ref59}. The strong summarization capability of LLMs enables detailed feedback generation that coherently summarizes students’ performance, which achieves a high agreement with the instructor when assessing the topic of students’ assignments \cite{ref5}. At the same time, ChatGPT can provide virtual assistance to medical school students to prepare for medical professional examinations \cite{ref37}. To aid in learning retention of important medical concepts, GPT-3.5-Turbo has demonstrated its ability to extract structured medical information from unstructured medical curriculum text \cite{ref60}. In our view, adapting the powerful open-source LLMs with medical-specific data allows an efficient alignment between LLMs and medical students on downstream tasks. The LLaMA-based MedAlpaca model could support medical students’ education through question-answering interactions to reinforce their knowledge and clarify lecture uncertainties \cite{ref60}.

\subsection{Healthcare Payers}
Healthcare insurance companies are in demand to accelerate patient insurance authorization by understanding patient information from a large number of electronic health records (EHR) datasets containing patient screening, treatment history, and outcomes. The document analytical capability of LLM allows the payers to accelerate the process of understanding unstructured clinical records, patient status, and insurance authorization. For instance, ChatGPT opens the possibility of writing pre-authorization requests for insurance companies \cite{ref61}. Also, payers could integrate LLMs into routine written communications and record keeping, leading to an increasing efficiency in tedious insurance administrative tasks. 

Sufficient knowledge alignment becomes a necessary foundation for safely applying LLM to patient clinical record understanding because the unique syntax and grammar in clinical notes are different from books or encyclopedias. GatorTron \cite{ref7} uses 90 billion words of text (including $>$82 billion words of de-identified clinical text) to align with payers' needs in extracting and interpreting patient characteristics from longitudinal EHRs. Such patient information documentation capabilities can also accelerate healthcare insurance companies' decision-making processes. Different from the non-generative BERT-based encoder architecture in GatorTron, GatorTronGPT \cite{ref6} is proposed as a generative GPT-based LLM, which is trained on the same de-identified clinical text. GatorTronGPT achieved superior performance on clinical-related tasks while also generating large-scale synthetic clinical text dataset (e.g., 20 billion words). The contribution of synthetic data offers a path to address the key challenge of data quantity and privacy. Comparing the performance of GatorTron trained on the synthetic and real-world clinical text datasets, synthetic clinical records promise to boost model performance by increasing the data diversity while protecting patient privacy.

Domain-specific model pretraining has shown a trustworthy performance in aligning LLM with clinical knowledge from large-scale, real-world data. Yet not all healthcare payers have sufficient clinical data and computing resources to develop their in-house LLM. In our view, common finetuning techniques will continue to play an important role in enabling LLMs to acquire expertise in understanding patient information. Aligning with the patterns in clinical notes, the task-specific finetuning enables ClinicalBERT \cite{ref8} to achieve good performance in predicting 30-day hospital readmission at various time points of admission, including early stages and discharge. Furthermore, using LLM to manage timely updates on consumers’ health status can be considered to simplify healthcare insurance administration. Foresight \cite{ref62} is an example of using named entity recognition and linking tools to convert EHR document text into structured concepts, enabling probabilistic forecasts for future medical events, such as disorders, substances, procedures, and findings.

\section{Outlook}
In the era of human-LLM collaboration, emerging efforts have attempted to align LLMs’ behavior with human knowledge, expectations, and value \cite{ref17,ref9}. Several priorities are required to improve the alignment performance towards a trustworthy interaction in various healthcare workflows. We discuss the challenge and outlook toward better aligning LLMs with healthcare stakeholders.

The human-AI alignment should carefully consider the linguistic, cultural, and educational backgrounds of the population worldwide. The linguistic generalizability is a key indicator of reliable alignment between LLMs and healthcare stakeholders. Compared with non-English speaking populations, current LLMs are easier to align with English-speaking healthcare stakeholders because of the training data source. ChatGPT is found to be able to pass the medical licensing examinations in English \cite{ref36,ref37,ref38} while failing in Asian languages \cite{ref63,ref64}. Extending the diversity of language in the training corpora becomes necessary, enabling the mitigation of linguistic discrepancies among the human-LLM alignment. Meanwhile, the integration of multimodal clinical data (e.g., wearable biosensor data, ambient sensor data, and omics data) is needed in a wider range of health demands. For example, handling audio data could open the possibility for promoting LLMs to meet the demand of ambient clinical documentation \cite{ref65}, allowing disabled patients to reduce the burden of communication and documentation workflows.

Reducing the misalignment between general- and healthcare-domain knowledge is highly dependent on human efforts in collecting healthcare-specific data, designing clear instruction prompts, and providing feedback for guiding LLMs \cite{ref75,ref76}. To date, it remains costly to involve human experts in curating a comprehensive dataset or providing detailed feedback to each output of LLMs. Alternatively, AI-generated data have promising potential to expedite LLM training and alignment to human preferences. Synthetic data generation can produce large-scale, privacy-free healthcare datasets to augment the lack of real-world data for LLM pre- and intermediate-training \cite{ref66}. In the post-training stage, the conventional nature of RLHF highly relies on time-intensive human-generated preference feedback, which is challenging to implement across all model outputs. We expect that AI-generated data can help scale RLHF with immediate AI feedbacks \cite{ref67}. It is thus a faster process of RL from AI Feedback (RLAIF) \cite{ref67} that combines human feedback with AI-generated feedback for training specialized models in the healthcare system.

To ensure a trustworthy alignment, we also need to avoid the abused use of LLM-generated data, such as promoting unfactual knowledge and misleading drug development \cite{ref68}. LLMs have been proven capable of generating novel insights for disease understanding \cite{ref44}, which can be a double-edged sword for healthcare stakeholders. Despite the promise of generating insights, these outcomes were unverified and unaligned with human knowledge, which is known as model hallucination \cite{ref69}. LLMs can be simply manipulated by harmful patterns or information in the training data and produce unfactual information, which will mislead patients and junior students in medical school. Hence, expert-verified training data becomes vital for safely aligning LLMs with healthcare stakeholders. Substantial efforts will be made to provide human feedback toward the quality improvement of LLM outputs. Alternatively, the LLM-generated answer-detection models (e.g., GPTZero \cite{ref70} and DetectGPT \cite{ref71}) can help regulate answer misuse and nonfactual information popularization.

Regulatory criteria are essential for the safe use of AI, ensuring that LLMs align with the expectations, values, and legal standards of healthcare stakeholders. In both the EU and the USA, all clinical decision support software for patients and healthcare professionals must undergo a medical device registration and approval process from regulators, such as the U.S. Food and Drug Administration (FDA) \cite{ref72}. Nowadays, the lack of systematic regulation standards becomes an obstacle to trustworthy interactions between healthcare stakeholders and LLMs \cite{ref73}. FDA has proposed a total product lifecycle (TPLC) approach to regulating AI/ML-based Software as a Medical Device (SaMD), which focuses on the continuous monitoring and improvement of these technologies throughout their lifespan \cite{ref74}. Inspired by SaMD, several directions of LLM regulatory standards could be considered for ensuring trustworthy human-LLM alignments \cite{ref15}: (1) well-defining the purpose and utility scope of LLMs in healthcare, (2) evaluating LLMs on the authoritative clinical data, and (3) releasing the easy-understanding stakeholder usage guideline. These protocols can improve the transparency and reliability of the human-LLM alignment. It is also urged for the open-source community to release and share LLM variations and development details with the healthcare community. We expect to see more systematic details on LLM model architecture, training parameters setting, and data distribution. Inspired by the open-source software development platform (e.g., Github) \cite{ref77}, the permission to control and update LLM versions can ensure up-to-date healthcare knowledge embedding and user feedback collection.

\section{Conclusion}
In the era of human-AI collaboration, the alignment between large language models (LLMs) and healthcare stakeholders opens up remarkable opportunities for productivity improvement in knowledge acquisition, workload reduction, and acceleration of clinical workflows. However, current healthcare systems are not yet fully prepared to adopt this enabling technology. Major efforts to align LLMs’ behavior with the knowledge and values of healthcare stakeholders across diverse scenarios are only beginning to emerge. Human involvement and feedback will continue to play a vital role during both the pre-training and post-training stages of LLM development. Evolving challenges remain to be addressed to ensure safe, robust, and trustworthy alignments between LLMs and healthcare stakeholders, including the implementation of appropriate regulatory frameworks to support the broader adoption of this technology within the healthcare community.

\bibliographystyle{unsrtnat}
\bibliography{references}  

\end{document}